\documentclass[]{interact}

\usepackage[numbers,sort&compress,merge]{natbib}% Citation support using natbib.sty
\bibpunct[, ]{[}{]}{,}{n}{,}{,}% Citation support using natbib.sty
\usepackage{physics} % Provides important physics notations. 

\newcommand{\ie}{i.e.}

\usepackage{xcolor}

\begin{document}

%\articletype{ARTICLE TEMPLATE}% Specify the article type or omit as appropriate

\title{Effect of neighbouring molecules on ground-state properties of many-body polar linear rotor systems}

\author{
\name{Tapas Sahoo\thanks{CONTACT Tapas Sahoo. Email: tsahoo@bose.res.in} and Gautam Gangopadhyay}
\affil{S. N. Bose National Centre for Basic Sciences, Salt Lake, Kolkata 700106, India}
}

\maketitle

\begin{abstract}
A path integral ground state approach has been used to estimate the ground-state energy and structural properties of hydrogen fluoride molecules pinned to a one-dimensional lattice. In the simulations, the molecules are assumed to be rigid, and only the continuous rotational degrees of freedom are considered. The constituents of a many-body system interact through the dipole-dipole interaction because the molecules have a permanent dipole moment. The workability of our approach has been demonstrated by estimating the ground-state energy, order parameter and nearest neighbour correlation for the systems of 2 and 3 HF molecules using quantum Monte Carlo simulations based on the path integral ground state methodology. The results agree satisfactorily with those obtained from exact Hamiltonian matrix diagonalization. In addition, the effect of neighbours on the ground state properties has been investigated for larger systems. The converged ground-state energy per neighbour as a function of inter-nuclear separation is considered to be the equation of state for the system. The chemical potential of the rotor chains also supports the convergence test.
\end{abstract}


\section{Introduction}
\label{sec:introduction}

In recent decades, scientists have focused on studying confined molecular systems in quasi-one-dimension, which have plenty of applications in biological systems. The confinement of molecules inside a Single-Walled Carbon Nanotube (SWCNT) is considered to be a suitable model to mimic various biological systems. It is possible to set the diameter of the SWCNT to be small enough to prevent entrapped molecules from passing through, making the system quasi one-dimensional. It has been reported~\cite{ma2017quasiphase,biskupek2020bond,chakraborty2019bonding} that confined molecules exhibit different structural, stability, reactivity, bonding, interactions, and dynamics effects. Therefore, confined molecular systems must be treated using the postulates of quantum mechanics because quantum phenomena are caused by confinement. Htoon and co-workers~\cite{ma2017quasiphase} recently conducted temperature-dependent (4.2~K to room temperature) photoluminescence spectroscopy on a one-dimensional single-file chain of water molecules encapsulated in (6,5) SWCNT. The orientational probability distributions revealed quasi-phase transitions between different arrangements of the entrapped water molecules, such as hydrogen-bonding, ferroelectric ordered, and disordered. Motivated by the experiments~\cite{ma2017quasiphase,biskupek2020bond} conducted for confined molecular systems, we have modelled the system of many-body rotor systems to compute the ground state properties of such systems. The purpose of this study is to investigate the effect of the number of neighbours on the ground state properties of non-ideal many-body rotor systems in which HF molecules are pinned to a one-dimensional lattice. When temperatures are low, the nuclear quantum effect plays a critical role in predicting properties of a system. But the classical molecular dynamics (MD) simulations fail to describe a system at very low temperature because the system turns into a ``quenched system'' without molecular motion in MD simulations. Hence, we need to study such a complex confined system using quantum molecular dynamics simulations.

When solving the time-dependent or independent Schr\"{o}dinger equation for a non-ideal system, it is often convenient to construct the wavefunction as a product of known basis functions. If there are $D$ degrees of freedom (DOFs) and each one requires $n$ basis functions, the size of the array required to store the wave function will be $n^D$, and for the Hamiltonian matrix, $n^{2D}$. Consequently, the computational cost of the exact basis set methods increases exponentially with the number of DOFs. This significantly oversized memory makes the exact methods impractical. However, the Density Matrix Renormalization Group (DMRG)~\cite{white1992density,white1993density,schollwock2005density,schollwock2011density} method is most effective for a many-body quantum problem when the system is of low dimension (in general) and the interaction energy is pair-wise additive. Recently, theoretical investigations of such confined systems of up to 100 linear rotors have been carried out employing the DMRG~\cite{iouchtchenko2018ground} and Multi-Layer Multi-Configuration Time-Dependent Hartree~\cite{mainali2021comparison} methods. In addition, the DMRG approach~\cite{serwatka2022ground,serwatka2022ferroelectric,serwatka2023quantum,serwatka2023optimized,serwatka2023quantum1} has been extended for the confined asymmetric top rotors pinned to a one-dimensional chain. The quantum phase transition was witnessed in both studies as entanglement entropy. However, DMRG will not be the automatic choice if the interaction between the molecules is not considered to be a pairwise additive. An alternative approach for the simulation of many-body non-ideal systems is the Path integral Monte Carlo methodology~\cite{ceperley1995path}, which is capable of simulating systems with any form of interaction. The arbitrary size of the system can be simulated. Indeed, the Path Integral approaches are suitable for studying quantum many-body systems, since the computational cost of this Monte Carlo-based method~\cite{metropolis1953equation} is low compared to the other methods.

Our primary objective is to evaluate the physical properties of the systems of HF molecules at the ground state. The well-known Path Integral Ground State (PIGS) approach~\cite{sarsa2000path,yan2017path} is used for such simulations, albeit the interaction between the rotors is pair-wise additive. The PIGS is a variational and computationally efficient method for estimating the energetic and structural properties of linear rotors with dipole interactions~\cite{abolins2011ground,abolins2013erratum,abolins2018quantum,sahoo2020path,sahoo2021path}. The method has recently been extended to compute the R\'{e}nyi entanglement entropy of a many-body rotor system using the replica trick~\cite{del2014particle,herdman2014path,herdman2014particle,herdman2017entanglement,sahoo2020path,sahoo2021path}.

The report includes a theoretical background including the Hamiltonian for the model system of HF molecules confined in one-dimensional chains, a pair-wise additive dipole-dipole interaction potential between two polar molecules, and a brief formulation of the PIGS approach for the linear rotor system in \textbf{Section}~\ref{sec:pigs}. The results are discussed in \textbf{Section}~\ref{sec:results}. The concluding remarks of this study are provided in \textbf{Section}~\ref{sec:conclusion}. 

\section{Theoretical background}
\label{sec:pigs}

Even though the PIGS methodology is discussed elsewhere~\cite{sarsa2000path,abolins2011ground,abolins2013erratum,yan2017path,abolins2018quantum,sahoo2020path,sahoo2021path}, the derivation is presented here to provide a complete understanding of this approach for the linear polar rotors.

\subsection{Hamiltonian and interaction potential}
\label{subsec:hamiltonian-and-potential}

\begin{figure}[htbp]
\centering
\includegraphics{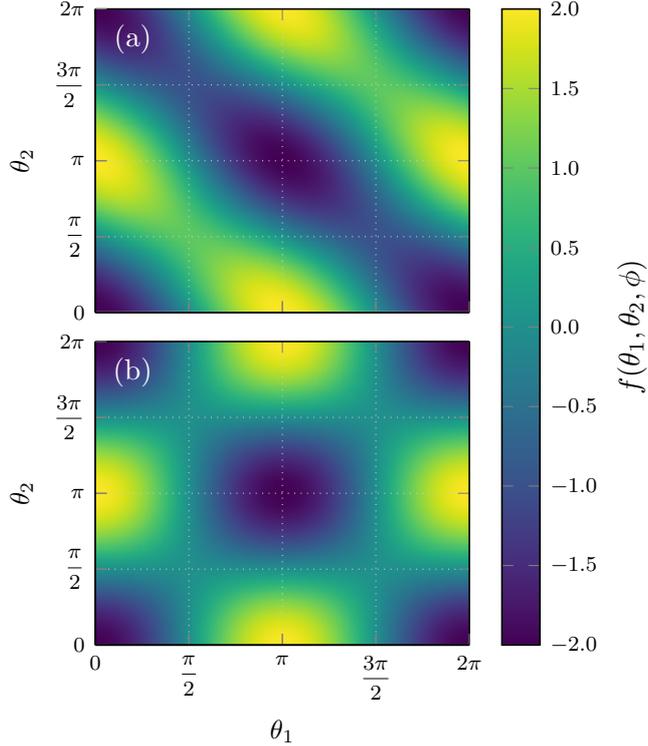}
\caption{Generic potential function $f\pqty{\theta_1,\theta_2,\phi}$ for the specific values of the azimuthal angle $\phi$ = 0 and $\dfrac{\pi}{2}$ is shown in panels (a) and (b), respectively.}
\label{fig:dipole-dipole-interaction-generic-function}
\end{figure}

Suppose that $N$ number of identical rigid linear rotors are pinned to a one-dimensional lattice. For such systems, the Hamiltonian operator is
\begin{align}
\hat{\mathcal{H}} = B\sum_{i=1}^N \hat{\ell}_i^2+\sum_{i=1}^{N-1}\sum_{j>i}^N\hat{V}_{ij},
\label{eq:linear-rotor-hamiltonian}
\end{align}
where $B$ is the rotational constant and $\hat{\ell}^2_i$ is the angular momentum operator of the $i^{\mathrm{th}}$ rotor. The second term on the left-hand side of Eq.~\eqref{eq:linear-rotor-hamiltonian} refers to the dipole-dipole interaction potential energy operator. In fact, molecular interactions are considered to be pair-wise and additive. If the rotors are placed along the $z$ axis in the space-fixed frame, in the position representation, the pair-wise interaction term is transformed into:
\begin{align}
V_{ij}\pqty{\theta_i,\phi_i,\theta_j,\phi_j}&=\frac{1}{4\pi\varepsilon_{0}}\frac{\mu^{2}}{r_{ij}^{3}}[\sin(\theta_{i})\sin(\theta_{j})\cos(\phi)-2\cos(\theta_{i})\cos(\theta_{j})] \nonumber \\
&=\frac{1}{4\pi\varepsilon_{0}}\frac{\mu^{2}}{r_{ij}^{3}}\times f(\theta_i,\theta_j,\phi),
\label{eq:dipole-dipole-interaction-zdir}
\end{align}
where $r_{ij}$ is the intermolecular distance between the $i^{\mathrm{th}}$ and the $j^{\mathrm{th}}$ rotors. The quantities $\mu$ and $\epsilon_0$ are the dipole moment of a rigid rotor and the permittivity of the vacuum, respectively. The polar angles, $\theta_{i} \in [0,\pi] $ and $\phi_{i} \in [0,2\pi]$, describe the orientation of the $i^{\mathrm{th}}$ rotor in the space-fixed frame. The $\phi$ in the above equation is $\phi_i-\phi_j$. The shape of the pair-wise interaction potential energy surface is determined by the angular function $f(\theta_i,\theta_j,\phi)$ defined in Eq.~\eqref{eq:dipole-dipole-interaction-zdir}. The trigonometric form of the angular function clearly indicates that
\begin{align}
	-2\le f(\theta_i,\theta_j,\phi)\le 2.
\label{eq:dipole-dipole-interaction-boundary}
\end{align}
In Fig.~\ref{fig:dipole-dipole-interaction-generic-function}, the angular function is plotted for two values of the azimuthal angle $\phi$ = 0 and $\dfrac{\pi}{2}$. For both of these cases, it meets the constraint provided in Eq.~\eqref{eq:dipole-dipole-interaction-boundary}.

% Figures for the potential energy functions

\subsection{Path integral ground state approach}
\label{subsec:pigs-formulation}

Let $\omega_i = \pqty{\theta_i, \phi_i}$ be the overall orientation of the $i^{\mathrm{th}}$ rotor; therefore, the collective angular motion of the system of $N$ rotors can be defined as $\mathbf{\Omega} = (\omega_{1}, \omega_{2}, \dots, \omega_{N})$. In PIGS approach, the ground-state ket $\ket{0}$ is approximated as 
\begin{align}
\ket{0}\propto\lim_{\beta\to\infty}e^{-\frac{\beta}{2}\hat{\mathcal{H}}}\ket{\Psi_\mathrm{T}},
\label{eq:pigs-ground-state-ket}
\end{align}
where the inverse temperature $\beta=\frac{1}{k_BT}$. $T$ is the absolute temperature and $k_B$ is the Boltzmann constant. The imaginary time projection operator $e^{-\frac{\beta}{2}\hat{\mathcal{H}}}$ relaxes the real-valued trial wavefunction $\ket{\Psi_\mathrm{T}}$ to the \emph{true} ground-state in the $\beta \to \infty$ limit. It should be noted that the approximation is valid only if the trial wavefunction has a non-zero overlap with the \emph{true} ground state, \ie, $\ip{0}{\Psi_{\mathrm{T}}}\neq0$. Moreover, as the PIGS is a variational method, the rate of the convergence of the limit in Eq.~\eqref{eq:pigs-ground-state-ket} must be affected by the choice of trial wavefunction. In the present simulations, a constant trial function, $\Psi_\mathrm{T}(\mathbf{\Omega}) = 1$ is chosen. This corresponds to the \emph{true} ground state of a set of non-interacting rotors. 

Using the ground-state ket defined in Eq. \eqref{eq:pigs-ground-state-ket}, we obtain the expectation value of an operator $\hat{O}$ as 
\begin{align}
\ev*{\hat{O}} = \lim_{\beta\to\infty}\frac{\ev*{e^{-\frac{\beta}{2}\hat{\mathcal{H}}}\hat{O}e^{-\frac{\beta}{2}\hat{\mathcal{H}}}}{\Psi_\mathrm{T}}}{\ev*{e^{-\beta\hat{\mathcal{H}}}}{\Psi_\mathrm{T}}}~
\label{eq:pigs-expectation-value}
\end{align}
\noindent
which in the position representation becomes  
\begin{align}
\ev*{\hat{O}} &= \frac{1}{Z_0}\lim_{\beta\to\infty}\int \dd{\mathbf{\Omega}} \int \dd{\mathbf{\Omega}^{\prime}} \Psi_\mathrm{T}\pqty{\mathbf{\Omega}} \mel*{\mathbf{\Omega}}{e^{-\frac{\beta}{2}\mathcal{\hat{H}}}\hat{O} e^{-\frac{\beta}{2}\hat{\mathcal{H}}}}{\mathbf{\Omega}^{\prime}}\Psi_\mathrm{T}\pqty{\mathbf{\Omega}^{\prime}},
\label{eq:expect_op_position}
\end{align}
where $Z_{0}$ is a normalizing pseudo-partition function of the form
\begin{align}
Z_{0} = \lim_{\beta\to\infty}\int \dd{\mathbf{\Omega}} \int \dd{\mathbf{\Omega}^{\prime}} \Psi_\mathrm{T}(\mathbf{\Omega})\mel*{\mathbf{\Omega}}{e^{-\beta\hat{\mathcal{H}}}}{\mathbf{\Omega}^{\prime}}\Psi_\mathrm{T}\pqty{\mathbf{\Omega}^{\prime}},
\label{eq:def_Z0}
\end{align}
and $\Psi_{\mathrm{T}}\pqty{\mathbf{\Omega}}=\ip{\mathbf{\Omega}}{\Psi_{\mathrm{T}}}$. 

The kinetic energy operator ($\hat{K}$) and the potential energy operator ($\hat{V}$) do not commute, \ie, $\comm*{\hat{K}}{\hat{V}}\neq 0$; therefore, the evaluation of the imaginary time propagator $\mel{\mathbf{\Omega}}{e^{-\beta\hat{\mathcal{H}}}}{\mathbf{\Omega}^{\prime}}$ and consequently, the computation of the pseudo-partition function become non-trivial problems. However, we are fortunate that the propagator can be evaluated successfully employing the Trotter factorization scheme~\cite{trotter1959product} stated as
\begin{align}
e^{-\tau\hat{\mathcal{H}}} = \lim_{\tau \rightarrow 0} e^{-\tau/2\hat{V}} e^{-\tau\hat{K}}e^{-\tau/2\hat{V}}.
\label{eq:trotter-fact}
\end{align}
To implement the above factorization scheme, the exponential operator in Eq.~\eqref{eq:def_Z0} is decomposed into $P-1$ discrete terms. Thus, the pseudo-partition function turns into 
\begin{align}
Z_{0} = \lim_{\substack{\beta\to\infty \\ \tau \to 0}} \int \dd{\mathbf{\Omega}} \int \dd{\mathbf{\Omega}^{\prime}} \Psi_\mathrm{T}\pqty{\mathbf{\Omega}}\mel*{\mathbf{\Omega}}{\pqty{\prod_{t=1}^{P-1}e^{-\tau\mathcal{\hat{H}}}}}{\mathbf{\Omega}^{\prime}}\Psi_\mathrm{T}\pqty{\mathbf{\Omega}^{\prime}},
\label{eq:split_Z0}
\end{align}
where $\tau = \beta/\pqty{P-1}$. The quantity $\tau$ is referred to as the imaginary time slice, and $P$ is the Trotter number, also called the quantum bead. Each quantum bead is numbered by $t$. The insertion of resolution of identity $\int \dd{x} \dyad{x} = \hat{1}$ between each pair of exponential operators, followed by the use of the Trotter factorization scheme~\cite{trotter1959product}, yields the following form of the partition function,
\begin{align}
Z_{0} &= \lim_{\substack{\beta\to\infty \\ \tau \to 0}} \int \dd{\mathbf{\Omega}_{1}} \cdots \int \dd{\mathbf{\Omega}_{P}} \Psi_\mathrm{T}\pqty{\mathbf{\Omega}_{1}} \Psi_\mathrm{T}\pqty{\mathbf{\Omega}_{P}} \nonumber \\ 
& \times \prod_{t=1}^{P-1} \exp\pqty{-\dfrac{\tau}{2} V\pqty{\mathbf{\Omega}_{t}}} \mel{\mathbf{\Omega}_{t}}{\exp\left(-\tau\hat{K}\right)}{\mathbf{\Omega}_{t+1}} \exp\pqty{-\dfrac{\tau}{2} V\pqty{\mathbf{\Omega}_{t+1}}} \nonumber \\
	&=\lim_{\substack{\beta\to\infty \\ \tau \to 0}} \int \dd{\mathbf{\Omega}_{1}} \cdots \int \dd{\mathbf{\Omega}_{P}} \Pi\pqty{\mathbf{\Omega}_{1}, \cdots,  \mathbf{\Omega}_{P}},
\label{eq:trotter_def_Z0}
\end{align}
where $\Pi(\mathbf{\Omega}_{1}, \cdots, \mathbf{\Omega}_{P})$ represents a multivariate distribution function of path variables. 

In the position representation, the high-temperature density matrix of a free rotor, also known as the rotational propagator, is off-diagonal. The many-body rotational propagator is a product of one-body factors,
\begin{align}
    \mel{\mathbf{\Omega}_{t}}{\exp\pqty{-\tau\hat{K}}}{\mathbf{\Omega}_{t+1}}
    = \prod_{i=1}^{N}\mel{\omega^{\bqty{t}}_{i}}{\exp\pqty{-\tau\hat{K}_i}}{\omega^{\bqty{t+1}}_{i}}
    = \prod_{i=1}^{N} \rho\pqty{\omega^{\bqty{t}}_{i}, \omega^{\bqty{t+1}}_{i}; \tau}.
	\label{eq:kin_dens}
\end{align}
The element of the high-temperature rotational density matrix~\cite{marx1999path} for the $i^{\mathrm{th}}$ free linear rotor is
\begin{align}
\rho\pqty{\omega^{\bqty{t}}_{i}, \omega^{\bqty{t+1}}_{i};\tau}=\sum_{l=0}^{\infty} \frac{2 l+1}{4 \pi} e^{-\tau B l\pqty{l+1}} P_{l}\pqty{\cos\pqty{\gamma^{\bqty{t}}_{i}}},
\end{align}
where
\begin{align}
\cos\pqty{\gamma^{\bqty{t}}_{i}} = \sin\pqty{\theta^{\bqty{t}}_{i}} \sin\pqty{\theta^{\bqty{t+1}}_{i}} \cos\pqty{\phi^{\bqty{t+1}}_{i}-\phi^{\bqty{t}}_{i}}
					&+\cos\pqty{\theta^{\bqty{t}}_{i}} \cos\pqty{\theta^{\bqty{t+1}}_{i}}
\end{align}
and $P_l\pqty{x}$ is the Legendre polynomial of degree $l$. To reduce the computational cost, the matrix elements of the rotational propagator are calculated at evenly spaced grid points of $\cos\pqty{\gamma^{\bqty{t}}_{i}}$ at the beginning of the simulations and stored them into a file. Then, the value of the propagator on a specific orientation of a rotor is determined from the pre-evaluated values using the linear interpolation algorithm. 

\begin{figure}
\centering
\includegraphics{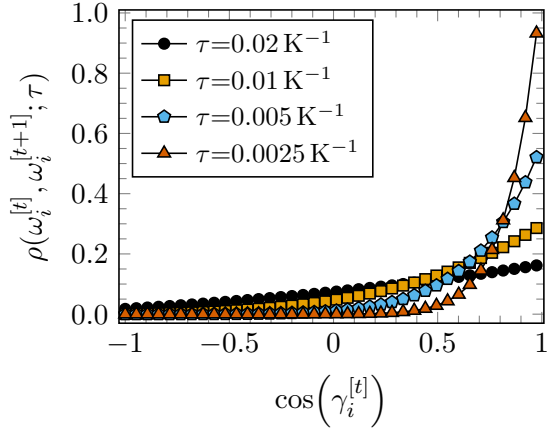}
\caption{An element of the high-temperature density matrix of a free HF molecule at fixed $\beta$ = 0.2 K$^{-1}$ is plotted for various $\tau$ values.}
\label{fig:rotational-propagator-linear-rotor}
\end{figure}

The prerequisite for Monte Carlo simulations~\cite{metropolis1953equation} is that the multivariate distribution function must be positive and real valued. The distribution function has two components: kinetic and potential. The potential contribution to the distribution function is always positive and real valued, since it is an exponential function (see Eq.~\eqref{eq:trotter_def_Z0}). In Fig. \ref{fig:rotational-propagator-linear-rotor}, an element of the high-temperature density matrix $\rho\pqty{\omega^{\bqty{t}}_{i}, \omega^{\bqty{t+1}}_{i};\tau}$ computed for a free HF molecule at fixed $\beta$ = 0.2 K$^{-1}$ is plotted as a function of $\cos\pqty{\gamma^{\bqty{t}}_{i}}$ for the quantum beads $P=10, 20, 40, 80$. The corresponding $\tau$ values are mentioned in the legend of the figure. The plots indicate that the rotational propagator is positive for every value of $\tau$. Therefore, the kinetic and potential terms are always positive, which means that the distribution function is always positive and has a real value throughout the simulation.

% Figures for rotational propagator

\begin{figure}
\centering
\includegraphics{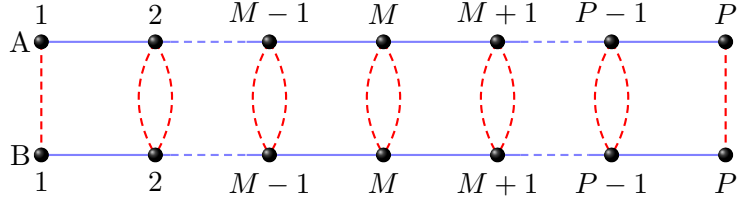}
\caption{Schematic diagram of $\Pi\pqty{\mathbf{\Omega}_{1}, \cdots,  \mathbf{\Omega}_{P}}$ for two rotors~\cite{Dmitri-thesis,rpqmc-1234-145-2016} $A$ and $B$ when the trial wavefunction, $\Psi_\mathrm{T}(\mathbf{\Omega}) = 1$. The black balls are assumed to be the quasi-particles originated from the Trotter factorization. They are numbered by a quantum bead or Trotter number, from 1 to $P$. The index $M$ denotes the middle bead. The solid blue lines represent the kinetic links $\rho\pqty{\omega^{\bqty{t}}_{i}, \omega^{\bqty{t+1}}_{i}; \tau}$ between the adjacent $t^{\text{th}}$ and $(t+1)^{\text{th}}$ beads, whereas the dashed green curve lines refer to the interaction potential, $\exp\pqty{-\frac{\tau}{2} V\pqty{\omega^{\bqty{t}}_{i}, \omega^{\bqty{t}}_{j}}}$, between quasi-particles, where $i,j \in \Bqty{A,B}$.}
\label{fig:path_explanation}
\end{figure}

A graphical illustration of the multivariate distribution function, $\Pi\pqty{\mathbf{\Omega}_{1}, \cdots, \mathbf{\Omega}_{P}}$, for a system of two rotors $A$ and $B$ is presented in Fig. \ref{fig:path_explanation}. The terminal beads are linked with a single potential link $\exp\pqty{-\frac{\tau}{2} V\pqty{\omega^{\bqty{t}}_{A}, \omega^{\bqty{t}}_{B}}}$ due to open nature of the imaginary time path. In the present work, we have utilized the Metropolis Monte Carlo algorithm~\cite{metropolis1953equation} to sample the orientations of the rotors according to the distribution function, $ \Pi(\mathbf{\Omega}_{1}, \cdots,  \mathbf{\Omega}_{P})$. Furthermore, Fig.~\ref{fig:path_explanation} shows that the Monte Carlo update of the overall orientation of a bead requires only the information of the adjacent beads. Specifically, only adjacent odd beads are required when the even beads get updated, or vice versa. Using this property of the distribution function, it is possible to simulate the odd and even beads separately based on shared memory parallelization algorithm~\cite{zeng2016moribs}.

\subsection{Expectation values in PIGS}
\label{subsec:expectation-value}

Using the same procedure as was used for the derivation of the pseudo-partition function, $Z_0$, we obtain the expression of the expectation value of an operator $\hat{A}$ that is diagonal in the position representation as
\begin{equation}
\ev{\hat{A}}{0}= \lim_{\substack{\beta\to\infty \\ \tau \to 0}} \dfrac{ \int \dd{\mathbf{\Omega}_{1}} \cdots \int \dd{\mathbf{\Omega}_{P}}  A\pqty{\mathbf{\Omega}_{P/2}} \Pi\pqty{\mathbf{\Omega}_{1}, \cdots,  \mathbf{\Omega}_{P}}}{ \int \dd{\mathbf{\Omega}_{1}} \cdots \int \dd{\mathbf{\Omega}_{P}} \Pi\pqty{\mathbf{\Omega}_{1}, \cdots,  \mathbf{\Omega}_{P}}}.
\label{eq:expect_op1}
\end{equation}
On the other hand, if an operator $\hat{B}$ is expressed as a function of the Hamiltonian operator, $\hat{\mathcal{H}}$, the operator $\hat{B}$ commutes the exponential operator, $e^{-\frac{\beta}{2}\mathcal{\hat{H}}}$, which establishes the relation $\hat{B} e^{-\frac{\beta}{2}\hat{\mathcal{H}}} = e^{-\frac{\beta}{2}\hat{\mathcal{H}}} \hat{B} $. Therefore, the numerator of the expectation value expression defined in Eq.~\eqref{eq:expect_op_position} becomes
\begin{align}
     \mel{\mathbf{\Omega}}{e^{-\frac{\beta}{2}\hat{\mathcal{H}}}\hat{B} e^{-\frac{\beta}{2}\hat{\mathcal{H}}}}{\mathbf{\Omega}^{\prime}}
     =\mel{\mathbf{\Omega}}{e^{-\beta\hat{\mathcal{H}}}\hat{B}}{\mathbf{\Omega}^{\prime}}
     =\mel{\mathbf{\Omega}}{\hat{B}e^{-\beta\hat{\mathcal{H}}}}{\mathbf{\Omega}^{\prime}}
\label{eq:o1_o2_commutator}
\end{align}
consequently, the expectation value of the $\hat{B}$ in turn becomes
\begin{equation}
\ev{\hat{B}}{0}= \lim_{\substack{\beta\to\infty \\ \tau \to 0}} \dfrac{ \int \dd{\mathbf{\Omega}_{1}} \cdots \int \dd{\mathbf{\Omega}_{P}}  \dfrac{\hat{B}\Psi_{T}\pqty{\mathbf{\Omega}_{k}}}{\Psi_{T}\pqty{\mathbf{\Omega}_{k}}} \Pi\pqty{\mathbf{\Omega}_{1}, \cdots,  \mathbf{\Omega}_{P}}}{ \int \dd{\mathbf{\Omega}_{1}} \cdots \int \dd{\mathbf{\Omega}_{P}} \Pi\pqty{\mathbf{\Omega}_{1}, \cdots,  \mathbf{\Omega}_{P}}},
\label{eq:expect_op2}
\end{equation}
where the index $k$ is either 1 or $P$ that means an equivalent expression exists for the first and the last beads. By replacing $\hat{B}$ with the total Hamiltonian operator, $\hat{\mathcal{H}}$, and using the action of the kinetic energy operator on a constant trial function, $\hat{K}\Psi_{T}(\mathbf{\Omega}_{P})$, that is zero, the expression of the total energy of the system at ground-state can be reduced to 
\begin{equation}
\ev{\hat{\mathcal{H}}}{0}=\lim_{\substack{\beta\to\infty \\ \tau \to 0}} \dfrac{ \int \dd{\mathbf{\Omega}_{1}} \cdots \int \dd{\mathbf{\Omega}_{P}}  V\pqty{\mathbf{\Omega}_{k}} \Pi\pqty{\mathbf{\Omega}_{1}, \cdots,  \mathbf{\Omega}_{P}}}{ \int \dd{\mathbf{\Omega}_{1}} \cdots \int \dd{\mathbf{\Omega}_{P}} \Pi\pqty{\mathbf{\Omega}_{1}, \cdots,  \mathbf{\Omega}_{P}}}.
\label{eq:expect_hamiltonian_op}
\end{equation}
Since both the limits, $\beta\to\infty$ and $\tau\to 0$, are impossible to achieve in practice, one can compute the expectation values as functions of $\beta$ and $\tau$. In \textbf{Section}~\ref{sec:results}, we have discussed how to determine both the limits in order to estimate the exact expectation values from the PIGS.

\subsection{Estimations of the expectation values using Monte Carlo simulation}
\label{subsec:estimator}

If we consider that the Markov process generates configurations $\mathbf{\Omega}_{\Bqty{1,\cdots,P}}^0,\mathbf{\Omega}_{\Bqty{1,\cdots,P}}^1, \cdots$, using the Metropolis algorithm~\cite{metropolis1953equation}, one can compute the properties associated with the structure as 
\begin{align}
\ev{\hat{A}}{0}\pqty{\tau;\beta}=\frac{1}{M}\sum_{i=i_{\text{first}}}^{i_{\text{last}}} A\pqty{\mathbf{\Omega}_{P/2}^i},
\label{eq:mc-estimator-for-structural-property}
\end{align}
and the total energy from
\begin{align}
E\pqty{\tau;\beta}=\frac{1}{M}\sum_{i=i_{\text{first}}}^{i_{\text{last}}} \frac{V\pqty{\mathbf{\Omega}_{1}^i)+V(\mathbf{\Omega}_{P}^i}}{2}.
\label{eq:mc-estimator-for-energy}
\end{align}
where $M=i_{\text{last}}-i_{\text{first}}+1$ is the number of steps in the average, $i_{\text{first}}$ is determined as the index of the simulation steps where the accumulated sum of the total energy gets equilibrated. It is important to pay attention to the positions of the beads used to estimate the physical properties. The structural properties of the system require only the orientation of the middle bead, whereas the terminal beads are involved in the estimation of the total energy. 

%expressed as the average of the contributions of two terminal beads, 1 and $P$ because Eq.~\eqref{eq:expect_op2} clearly shows that the contributions are equivalent to one another.

\section{Results and discussion}
\label{sec:results}
In this article, the effect of interaction strength on the ground-state properties of a series of many-body rotor systems composed of HF molecules pinned to a one-dimensional lattice has been investigated using PIGS approach. The study was restricted to continuous angular motions of HF molecules. The dipole-dipole interaction between the polar molecules is discussed in~\ref{subsec:hamiltonian-and-potential}. The rotational constant of the HF molecule is 20.559 cm$^{-1}$~\cite{rothschild1964pure} and the dipole moment is taken as 1.827 Debye~\cite{muenter1970hyperfine}. 

% Figures for the beta convergence
\begin{figure}
\centering
\includegraphics{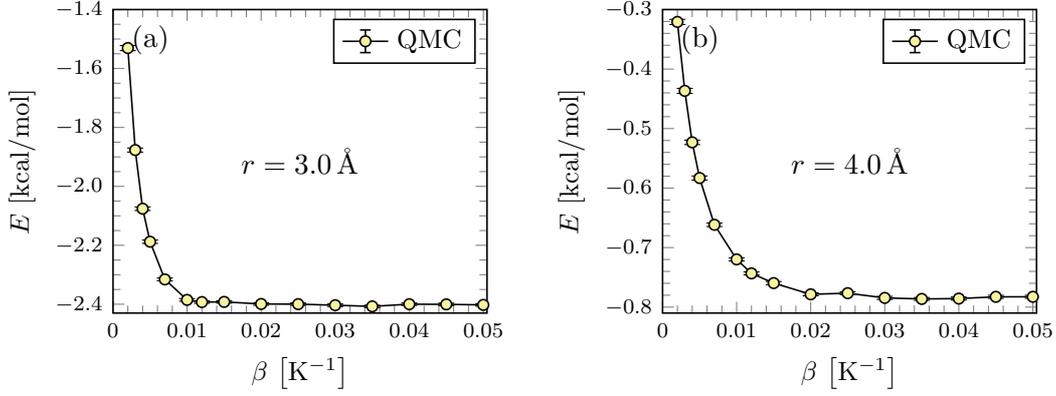}
\caption{The total energy, $E$, of two HF molecules at intermolecular separations of 3.0 and 4.0\AA~is shown in panels (a) and (b), respectively. The results have been obtained by employing the QMC simulations based on the multivariate distribution function expressed in Eq.~\eqref{eq:trotter_def_Z0} that is derived from the PIGS approach. The error bars are within the symbol size.}
\label{fig:total-energy-beta-convergence-n2}
\end{figure}

% tau convergence
\begin{figure}
\centering
\includegraphics{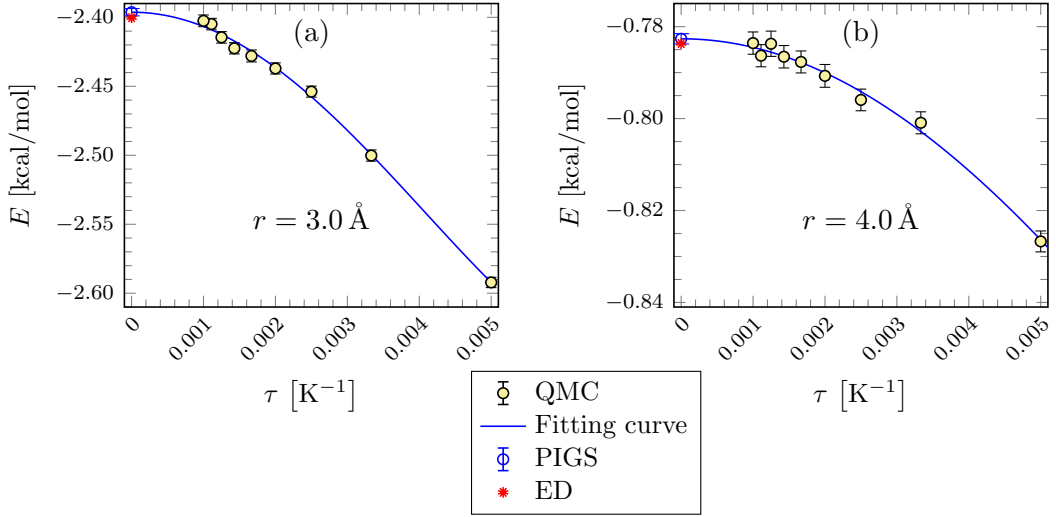}
\caption{The total energy, $E$, as a function of imaginary time, $\tau$, for the two coupled linear rotors at intermolecular separations of 3.0 and 4.0\AA~is shown in panels (a) and (b), respectively. A marker of a black circle filled with green color indicates the result estimated from the QMC simulation. The blue solid line represents the fitted curve. The markers of the blue circle with an error bar and the red 10-pointed star indicate the results of the PIGS and the ED simulations, respectively.}
\label{fig:total-energy-tau-convergence-n2}
\end{figure}

For intermolecular distances ($r$) of 3.0-10.0\AA,~the total energy $E$ of each of the rotor systems has been estimated using quantum Monte Carlo (QMC) simulation as a function of the total imaginary time length, $\beta$, at a particular time slice, $\tau=$0.005 K$^{-1}$. The simulations have been carried out based on the multivariate distribution function, $\Pi(\mathbf{\Omega}_{1}, \cdots, \mathbf{\Omega}_{P})$, defined in Eq.~\eqref{eq:trotter_def_Z0}. To estimate the total energies of the systems, we used Eq.~\eqref{eq:mc-estimator-for-energy} which only requires the angular positions of the terminal beads of each rotor. Figure \ref{fig:total-energy-beta-convergence-n2} shows the variation of the total energy of two HF molecules as a function of $\beta$ at two $r$ values. In panel (a), it is 3.0\AA, and in panel (b), it is 4.0\AA. The plots clearly indicate that the converged value of $\beta$ at these intermolecular separations can be estimated as 0.05 K$^{-1}$. We have also calculated the converged total imaginary time length for the other intermolecular distances. In order to ensure that the imaginary time length, $\beta$, is sufficiently long, its value has been fixed at 0.1 K$^{-1}$ for all the simulations performed for a set of Trotter numbers, $P$, for each intermolecular distance in this study.

% Figures for the ground-state energy
\begin{figure}
\centering
\includegraphics{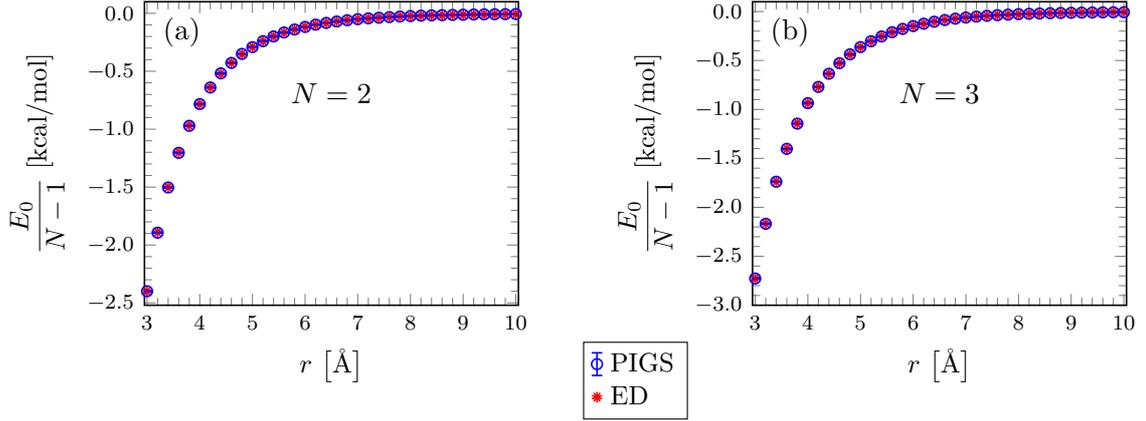}
\caption{The ground-state energy per number of neighbouring rotor, $\dfrac{E_0}{N-1}$, is displayed as a function of intermolecular distance, $r$, for the rotor systems with $N$=2 and 3 in panels (a) and (b), respectively. The results obtained from the ED and PIGS methodologies are marked by the red star and blue circle, respectively. The error bars associated with the PIGS results are within the symbol size.}
\label{fig:ground-state-energies-per-number-of-neighbouring-rotors-n2-3}
\end{figure}

% Figures for the order-parameter
\begin{figure}
\centering
\includegraphics{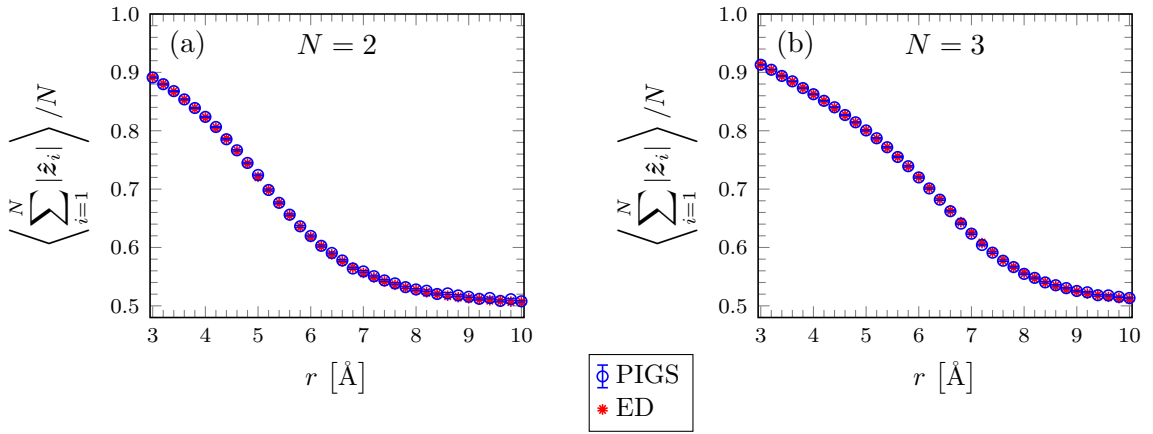}
\caption{In panels (a) and (b), the PIGS-simulated order parameter for each of the rotor systems with $N=2$ and 3 is compared to the order parameter obtained from the ED calculation as a function of $r$. The notations are the same as in Fig.~\ref{fig:ground-state-energies-per-number-of-neighbouring-rotors-n2-3}. The error bars associated with the PIGS results are within the symbol size.}
\label{fig:ground-state-order-parameter-for-rotors-n2-3}
\end{figure}

% Figures for the correlation function
\begin{figure}
\centering
\includegraphics{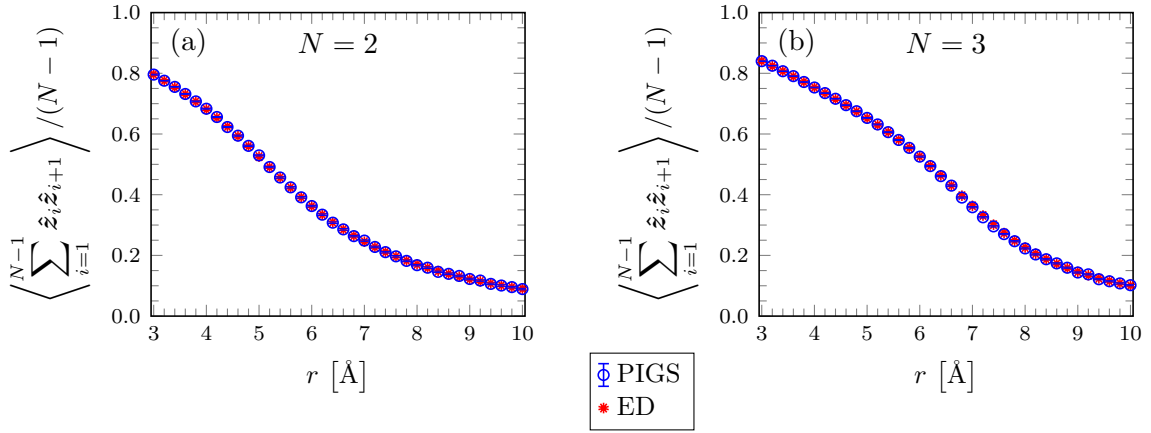}
\caption{The PIGS-simulated nearest neighbour correlation function for the rotor chains with $N$=2 and 3 is compared with the same obtained from the ED calculation as a function of $r$. The labels and notations of the figure are the same as in Fig.~\ref{fig:ground-state-energies-per-number-of-neighbouring-rotors-n2-3}. The error bars associated with the PIGS results are within the symbol size.}
\label{fig:ground-state-correlation-for-rotors-n2-3}
\end{figure}

In numerical calculations, the limit of $P\to\infty$ or $\tau\to 0$ is not possible to reach; therefore, the ground-state property of a many-body system can be obtained by extrapolating the results estimated from the QMC simulations at $\tau=0$. The total energy of each system has been calculated using the QMC simulation based on the PIGS approach as a function of $\tau$ at a fixed value of $\beta$=0.1 K$^{-1}$ for each intermolecular distance. Figure \ref{fig:total-energy-tau-convergence-n2} illustrates how the total energy of the system of two HF molecules depends on the imaginary time slices $\tau$ at the intermolecular distances 3.0 and 4.0\AA~in panels (a) and (b), respectively. To find the extrapolated data, we fit the QMC results to a quartic equation $E(\tau) = E_0+a\tau^2 + b\tau^4$. The quantities, $E_0$, $a$, $b$ are the fitting parameters. The ground-state energy $E_0$ of the system is found to be the extrapolated value of the data obtained from the fitting equation at the limit of $\tau=0$, i.e., $E_0=E(\tau=0)$. The ground-state energy computed from the PIGS method has an excellent accordance with the exact result obtained by diagonalizing the Hamiltonian matrix (ED). In the subsequent figures, the ``PIGS'' legend refers to the results obtained from the fitting curve at $\tau=0$ after fitting the QMC results. We have also used the same fitting procedure as for the $N=2$ system to estimate the ground-state energies and structural properties, such as order parameters and nearest neighbour correlations, for the larger systems at the intermolecular distances ranging from 3.0 to 10.0\AA. The order parameter and the nearest neighbour correlation are defined as $\expval{\displaystyle\sum_{i=1}^{N}\abs{\vu*{z}_i}}/N$ and $\expval{\displaystyle\sum_{i=1}^{N-1}\vu*{z}_i\vu*{z}_{i+1}}/(N-1) $, respectively, where $\vu*{z}_i$ represents the unit vector of the $i^{\mathrm{th}}$ rotor along the molecular axis in the center of mass coordinate system. Important to note that the order parameter and the nearest neighbour correlation are computed using Eq.~\eqref{eq:mc-estimator-for-structural-property} that needs only the polar angles at the middle bead of each rotor.

%In order to encounter the effect of neighbouring rotors on the ground-state properties of the rotor systems, 
To demonstrate the workability of PIGS approach, we have compared the ground-state energy per number of neighbour with the exact result as a function of intermolecular distance for the systems of $N=2$ and 3 in Fig. \ref{fig:ground-state-energies-per-number-of-neighbouring-rotors-n2-3}. In addition, the order parameter and nearest-neighbour correlation are compared with the exact results in Figs. \ref{fig:ground-state-order-parameter-for-rotors-n2-3} and \ref{fig:ground-state-correlation-for-rotors-n2-3}. The comparison between the PIGS results and the exact ones is satisfactory. The ground-state energy per number of neighbouring rotor decreases as the intermolecular distance decreases. The ground-state energy is lower when the molecules are getting closer together because the strength of the interaction between the molecules is inversely proportional to the distance between them (see Eq.~\eqref{eq:dipole-dipole-interaction-zdir}). The nearest neighbour correlation and the order parameter show an opposite trend compared to the ground-state energy of the systems. When the interaction strength increases, the polar molecules become more aligned, which increases the order parameter and nearest neighbour correlation.

% Plots for the convergence of ground-state equation of states for N=2-9 
\begin{figure}[htbp]
\centering
\includegraphics{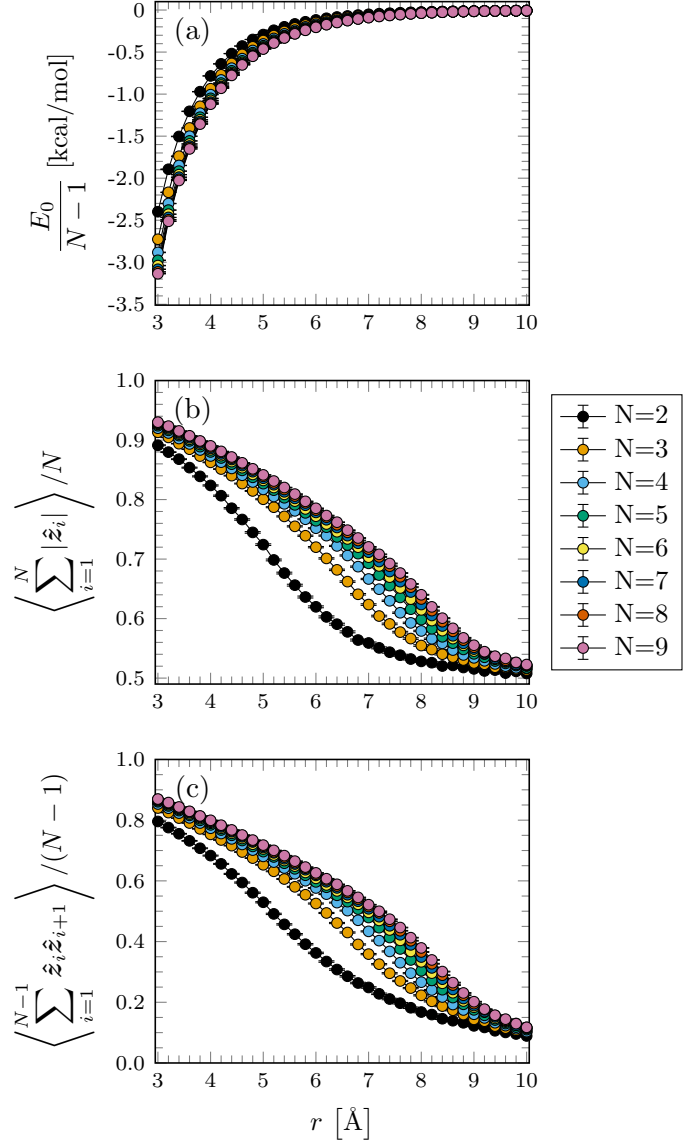}
\caption{The ground-state energy per number of rotor, order parameter, and nearest neighbour correlation of $N=2-9$ are shown in the panels (a-c) as a function of intermolecular distance. All the results are obtained from the PIGS simulations. The error bars are within the symbol size.}
\label{fig:eos-figure0}
\end{figure}

% Plots for the convergence of ground-state equation of states for N=10-17 
\begin{figure}[htbp]
\centering
\includegraphics{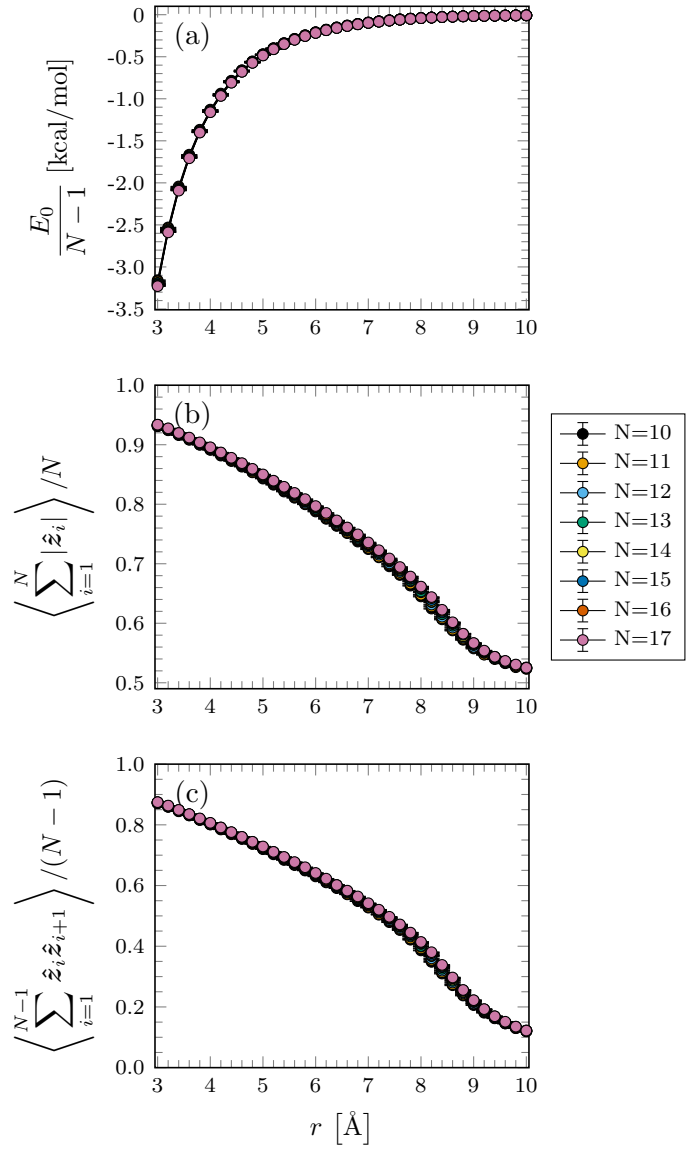}
\caption{Same as Fig.~\ref{fig:eos-figure0}, but for $N=10-17$. The error bars are within the symbol size.}
\label{fig:eos-figure1}
\end{figure}

% Plots for the convergence of ground-state equation of states for N=10-17 
\begin{figure}[htbp]
\centering
\includegraphics{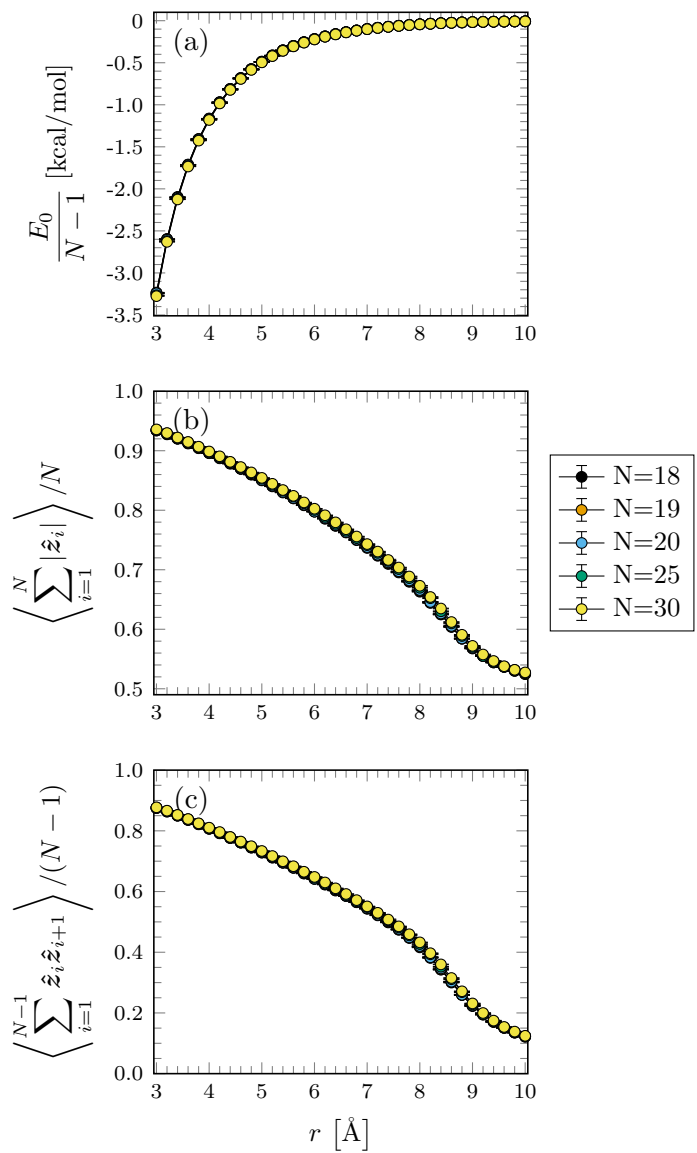}
\caption{Same as Fig.~\ref{fig:eos-figure0}, but for $N=18-20,25, 30$. The error bars are within the symbol size.}
\label{fig:eos-figure2}
\end{figure}

We have further examined the effect of neighbouring rotor on the ground-state properties of the many-body rotor systems. The ground-state energy per number of neighbouring rotors, order parameter, and nearest-neighbour correlation are presented in panels (a–c) of Figs.~\ref{fig:eos-figure0},~\ref{fig:eos-figure1}, and~\ref{fig:eos-figure2} for the systems up to $N=30$ as a function of intermolecular separation, $r$. Figures~\ref{fig:eos-figure1} and~\ref{fig:eos-figure2} demonstrate that the effect of neighbours on each ground-state property reaches the thermodynamic limit when the system has at least 25 rotors. The order parameter and the nearest-neighbor correlation exhibit rapid convergence in both low and high interaction regimes, in contrast to the intermediate interaction region, which is defined by intermolecular distances of approximately 6-9\AA. For one-dimensional lattice systems, the equation of state is determined by the effect of lattice length on the total energy of the system. The number of neighboring rotors and intermolecular separation can be used to express the lattice length. As a consequence, we can consider the converged ground-state energy per neighbouring rotor shown in panel (a) of Fig.~\ref{fig:eos-figure2} as the quantum equation of state for the many-body pinned rotors.

% Plots of ground-state chemical potential for r=3.0-4.8 angstrom
\begin{figure}[htbp]
\centering
\includegraphics{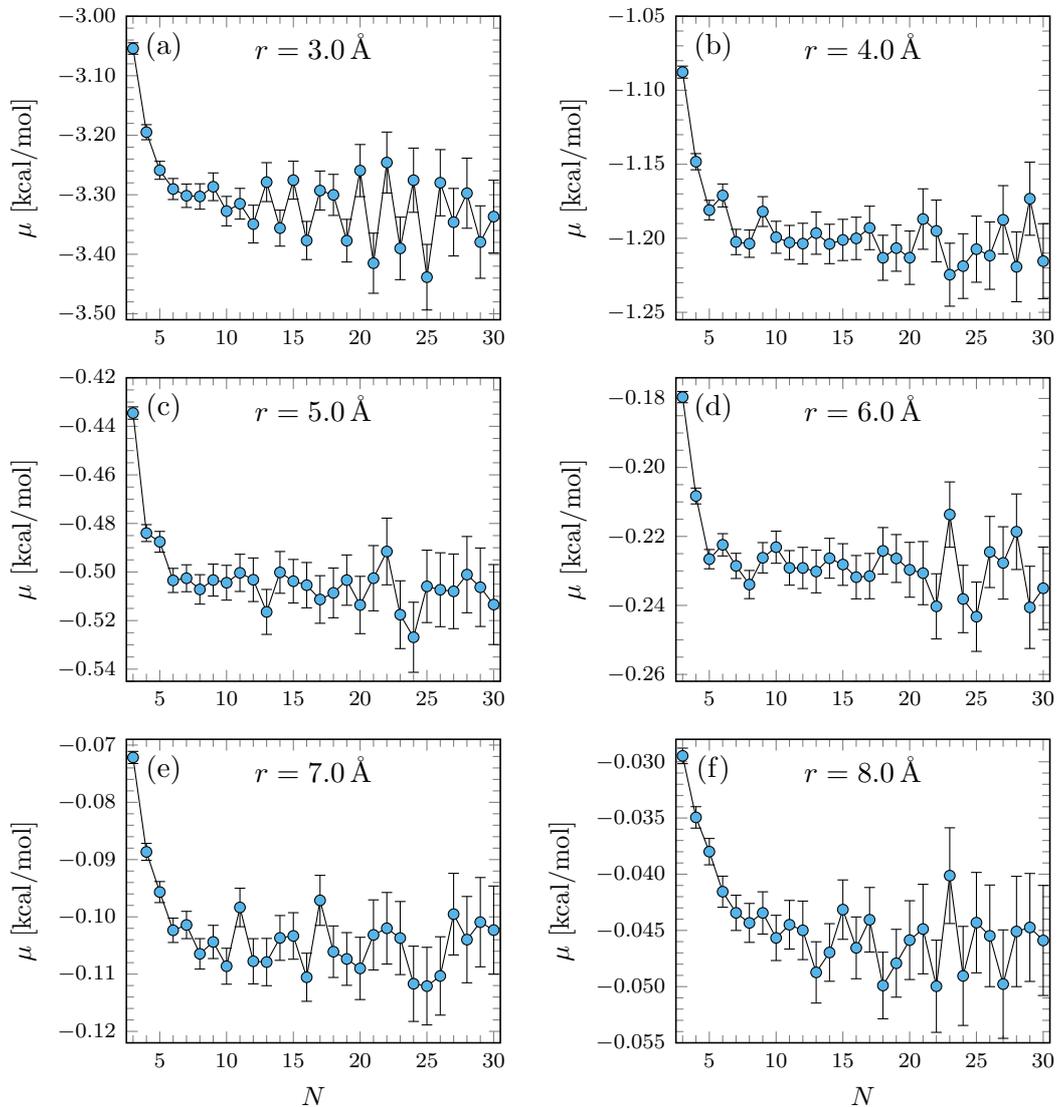}
\caption{The ground-state chemical potential $\mu$ simulated by the PIGS for the systems of 3-30 rotors is plotted for $r=$ 3.0, 4.0, 5.0, 6.0, 7.0 and 8.0\AA.}
\label{fig:chemical-potential-figure0}
\end{figure}

To ensure the convergence of the effect of neighbours on the energy of the rotor chains, we have also computed the ground-state chemical potential of the systems~\cite{halverson2018quantifying,iouchtchenko2018ground} and plotted it in Fig.~\ref{fig:chemical-potential-figure0} for the $r$ values of 3.0, 4.0, 5.0, 6.0, 7.0 and 8.0\AA. The chemical potential is defined as $\mu=E_0(N)-E_0(N-1)$, where $E_0(N)$ is the ground-state energy of $N$ rotors pinned to a one-dimensional chain. Each panel of Fig.~\ref{fig:chemical-potential-figure0} depicts that the chemical potential calculated at each inter-molecular distance decreases rapidly with the addition of a rotor to the system, up to the system size of 10 rotors. However, for larger systems with up to 30 rotors, the quantity exhibits a jagged pattern, in contrast to the smooth curve obtained from DMRG results~\cite{iouchtchenko2018ground}.

\section{Concluding remarks}
\label{sec:conclusion}

This study addresses the workability of the PIGS approach and discusses the effect of interaction strength on the energetic and structural properties of many-body rigid polar molecules pinned to a one-dimensional lattice. For the sake of simplicity, only the continuous angular motion of such molecules is considered. In this study, we assume that polar HF molecules interact with each other through a pair-wise additive dipole-dipole interaction potential. The shape of the multivariate distribution function used in the QMC simulations derived from the PIGS approach is an open path in position representation. The total energy is estimated using the angular positions of the terminal beads of each rotor, while the structural property is computed from the middle bead of each rotor.

The ground-state energy, order parameter and nearest neighbour correlation agree well with exact calculations for both $N = 2$ and $N = 3$. The shorter the intermolecular distance, the stronger the interaction, and the lower the total ground state energy of each system. However, the order parameter and the nearest neighbour correlation of each system show the opposite trend because stronger interaction aligns the polar molecules in head-and-tail orientations, making the high values of the order parameter and the nearest neighbour correlation. In addition, we have also checked how neighbour rotors affect the ground-state properties of the systems. In such linear chain systems of HF molecules, 25 neighbours are sufficient to achieve convergent results. The converged ground-state energy of a system with $N=25$ or higher can be considered to be the quantum equation of state of many-body quantum rotor systems. 

This investigation could be extended to other polar or non-polar molecules in the future. Further improvements could be achieved by using translational motions instead of rotational degrees of freedom. Additionally, the quasi-phase transition~\cite{ma2017quasiphase} for a quasi-one-dimensional system can be understood by calculating the orientational ordering at various temperatures. The distribution function for these QMC simulations is a closed path in the position representation, while it is open for the PIGS. We also want to use the \emph{replica trick} algorithm~\cite{herdman2014particle,sahoo2020path} to estimate the R\'{e}nyi entanglement entropy for various linear and nonlinear rotor systems. This will help us understand the quantum phase transition for confined many-body rotor systems.

\section*{Acknowledgement(s)}
Dr. Tapas Sahoo acknowledges the advanced postdoctoral research programme supported by S. N. Bose National Centre for Basic Sciences, India for his postdoctoral fellowship and Professor Pierre-Nicholas Roy, University of Waterloo, Canada for the high-performance computational facility.

\bibliographystyle{tfo}
\bibliography{manuscript.bib}

\end{document}